\documentclass[prl,twocolumn]{revtex4}
\usepackage{graphicx}

\def\in-situ{{\em in\ situ}}
\def\ex-situ{{\em ex\ situ}}

\begin{document}

\title{Epitaxial influence on the ferromagnetic semiconducotor EuO}

\author{N.J.C. Ingle and I.S. Elfimov}
\affiliation{Advanced Materials and Process Engineering Laboratory, University of British Columbia, Vancouver, BC, Canada}

\date{\today}

\begin{abstract}
From first principles calculations we investigate the electronic structure and the magnetic properties of EuO under hydrostatic and epitaxial forces.  There is a complex interdependence of the O $2p$ and Eu $4f$ and $5d$ bands on the magnetism in EuO, and decreasing lattice parameters is an ideal method to increase the Curie temperature, $T_c$.  Compared to hydrostatic pressure, the out-of-plane compensation that is available to epitaxial films influences this increase in $T_c$, although it is minimized by the small value of poisson's ratio for EuO.  We find the semiconducting gap closes at a 6\% in-plane lattice compression for epitaxy, at which point a significant conceptual change must occur in the  active exchange mechanisms. 

\end{abstract}
\maketitle

The desire to connect the electronic and magnetic properties of materials to enable new device characteristics has renewed focus on the many unique properties of EuO, one of the initially discovered ferromagnetic semiconductors\cite{matthias-prl-61}.  The divalent Eu ions in EuO possess a very large local moment from the half filled $4f$ band producing a saturation magnetization of $7\mu_b$\cite{matthias-prl-61, mcguire-jap-64} while a gap of 1.2eV separates the half filled $4f$ band from the $5d6s$ conduction band.\cite{mauger-pr-86}  Low levels of electron doping are readily achieved with oxygen vacancies, EuO$_{1-x}$, leading to an insulator-to-metal transition (IMT) in conjunction with the ferromagnetic ordering\cite{shapira-prb-73a} and 100\% spin polarization of the conduction electrons.\cite{steeneken-prl-02}  EuO may therefore be a good candidate for a spin injection material.   Cation and anion doping has been shown to increase the Curie temperature ($T_c$) up to 170K,\cite{ott-prb-06,konno-jjap-96,schmehl-naturematerials-07}   from that of 69K for stoicheometric EuO.\cite{mcguire-jap-64}

The integration of EuO$_{1-x}$ with Si and GaN has been successfully demonstrated\cite{lettieri-apl-03,schmehl-naturematerials-07} so there is renewed interest in increasing $T_c$ towards room temperature to enable spintronic applications.  Beyond the effects of doping on $T_c$, it has been shown that a hydrostatic pressure of about 100 kbar can increase $T_c$ to 200 K.\cite{dimarzio-prb-87}   Although hydrostatic pressure is not an option for device applications, a similar effect on $T_c$ may be obtainable by using epitaxial strain.  However, epitaxy generates a biaxial stress state which is quite different from the isotropic stress state of hydrostatic pressure.  Furthermore, there are a number of competing exchange mechanisms that determine the magnetism in EuO,  and is is not known exactly how this competition plays out as the lattice parameters are changed.  

In this paper we explore the effects of biaxial stress in comparison with isotropic stress on the bandstructure and magnetism of EuO with density functional theory (DFT).  We find that the biaxial stress state expected in an epitaxial film will lead to a 50\% increase in the $T_c$ before an insulator to metal transition occurs.  This increase, although not as great as that found for isotropic stress, is much larger than expected for an 3-dimensional material under biaxial stress due to the small value of Poisson's ratio for EuO.   From a band structure perspective of the magnetism in EuO, we find that reducing lattice parameters is an ideal method of increasing $T_c$.   Although the closing of the semi-conducting gap, which we find at 5\% compression for isotropic stress and 6\% compression of biaxial stress, must lead to a conceptually different exchange mechanisms, no sudden changes are seen in the magnetic properties.  The combined effects of doping and epitaxy are explored and are found to generate a significant increase in the mean field $T_c$ of up to 175K.  

\begin{figure}
\includegraphics[width=0.47\textwidth]{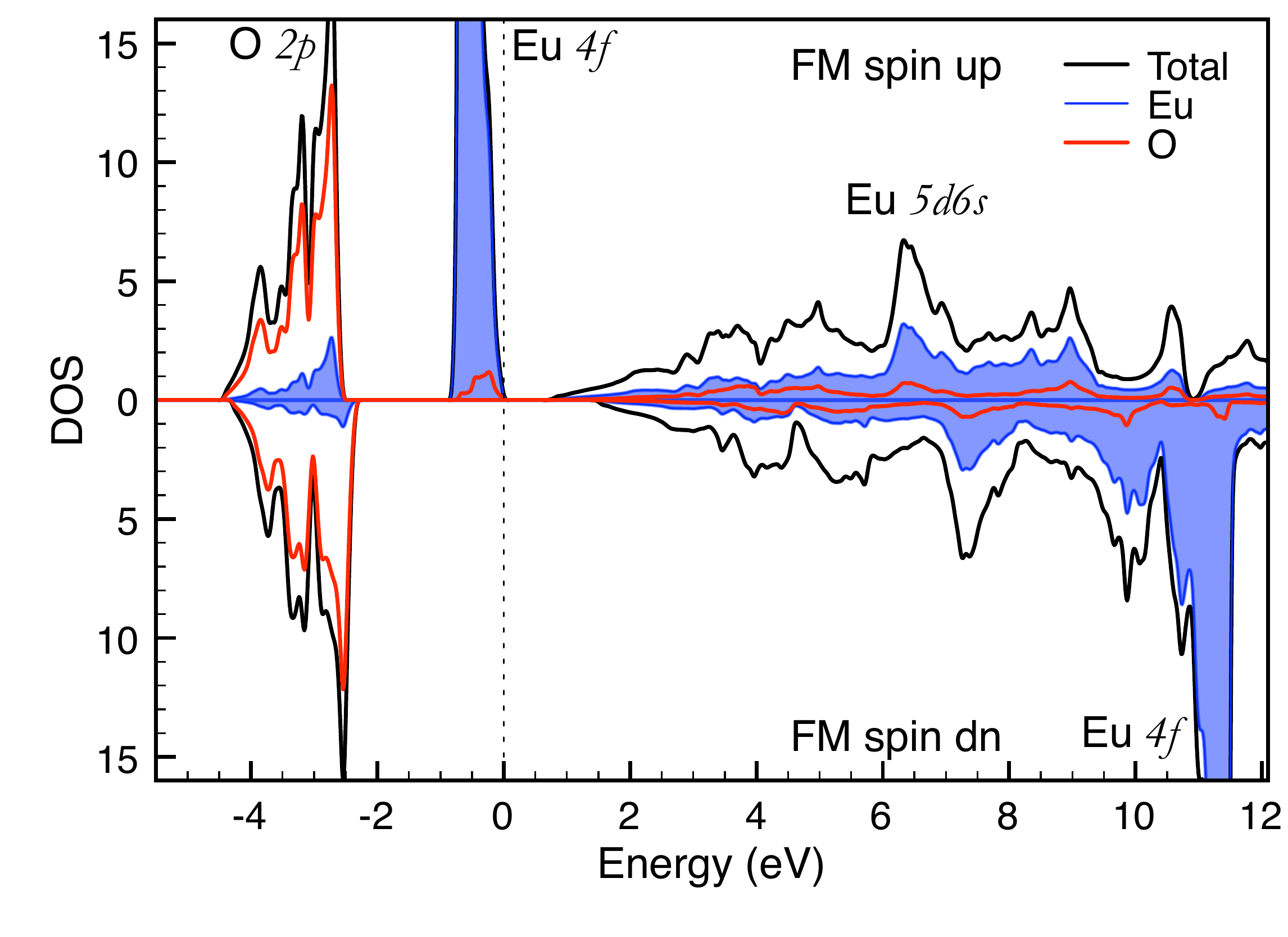}
\caption {(a) The spin resolved density of states of bulk EuO in the FM configuration with optimized bulk lattice parameters, a=5.1578\AA.  The bands are labeled and the zero of energy is at the Fermi energy, $E_f$.}
\label{FM-DOS-J}
\end{figure}

The EuO band structure is calculated with the full-potential linearized augmented plane-wave DFT code WIEN2K.\cite{blaha-01} The exchange and correlation effects are treated within the generalized gradient approximation (GGA), after Perdew {\it et al}.\cite{perdew-prl-96}  The LSDA+U method\cite{anisimov-prb-93} was used to account for strong correlations between the electrons in the Eu $4f$ shell. Note that standard GGA or local spin density approximation (LSDA) predict EuO to be a metal whereas measurements clearly show the existence of conductivity gap.\cite{mauger-pr-86}  The exchange parameter (J$_H$=0.77 eV) was chosen according to Ref. \onlinecite{van_der_marel-prb-88}, and the on-site Coulomb repulsion for the Eu $4f$  orbital was set to $U_f$=8.3eV, while that of the O $p$ orbital were $U_p = 4.6$eV and J$_H$=1.2 eV.\cite{slater-book-1960,ghijsen-prb-1988, altieri-thesis} With these values the LSDA+U band structure with the \{001\} antiferromagnetic spin configuration (AFM$_I$) or the NiO-type \{111\} AFM spin configuration (AFM$_{II}$) shows a gap of 1.2 and  1.3eV, respectively.  The ferromagnetic (FM) spin arrangement exhibits a gap of 0.7eV (see Figure \ref{FM-DOS-J}).  Both are consistent with the experimental optical abosorption gaps of 0.9 and 1.2 eV observed below and above the magnetic transition temperature.\cite{schoenes-prb-74,comment_on_para_vs_AFM}  Our calculations show a minimum in the total energy of the bulk EuO at a=5.1578\AA, in good agreement with the measured latticee parameter of 5.1439\AA.\cite{cunningham-rm-63}   

The calculations were done for the isotropic stress case of hydrostatic pressure, where the cubic unit cell is maintained, and the biaxial stress case of epitaxial film growth where the unit cell becomes tetragonal.  For all the biaxial stress case calculations, the in-plane lattice parameter is held fixed and the minimum in the total energy is found as a function of the $c$ lattice parameter.

Since the Eu 2+ ion is in the $f^7$ high spin configuration,  the magnetic properties of EuO are often described by a Heisenberg model.  According to the description of Mauger and Godart\cite{mauger-pr-86}, the nearest neighbor exchange, $J_1$, is based on a $4f$ electron having a virtual excitation to the $5d$ band where it experiences an exchange interaction with the $4f$ spin on a nearest neighbor. This leads to ferromagnetic coupling.  The nature of the next nearest neighbor exchange constant, $J_2$, is thought to be rather complex and involves various competing exchange paths between the O 2$p$, Eu 5$d$ and 4$f$ orbitals. A model calculation of $J_2$ in EuO shows that it is still ferromagnetic, but only about 30\% the size of $J_1$.\cite{mauger-pr-86}

\begin{figure}
\includegraphics[width=0.47\textwidth]{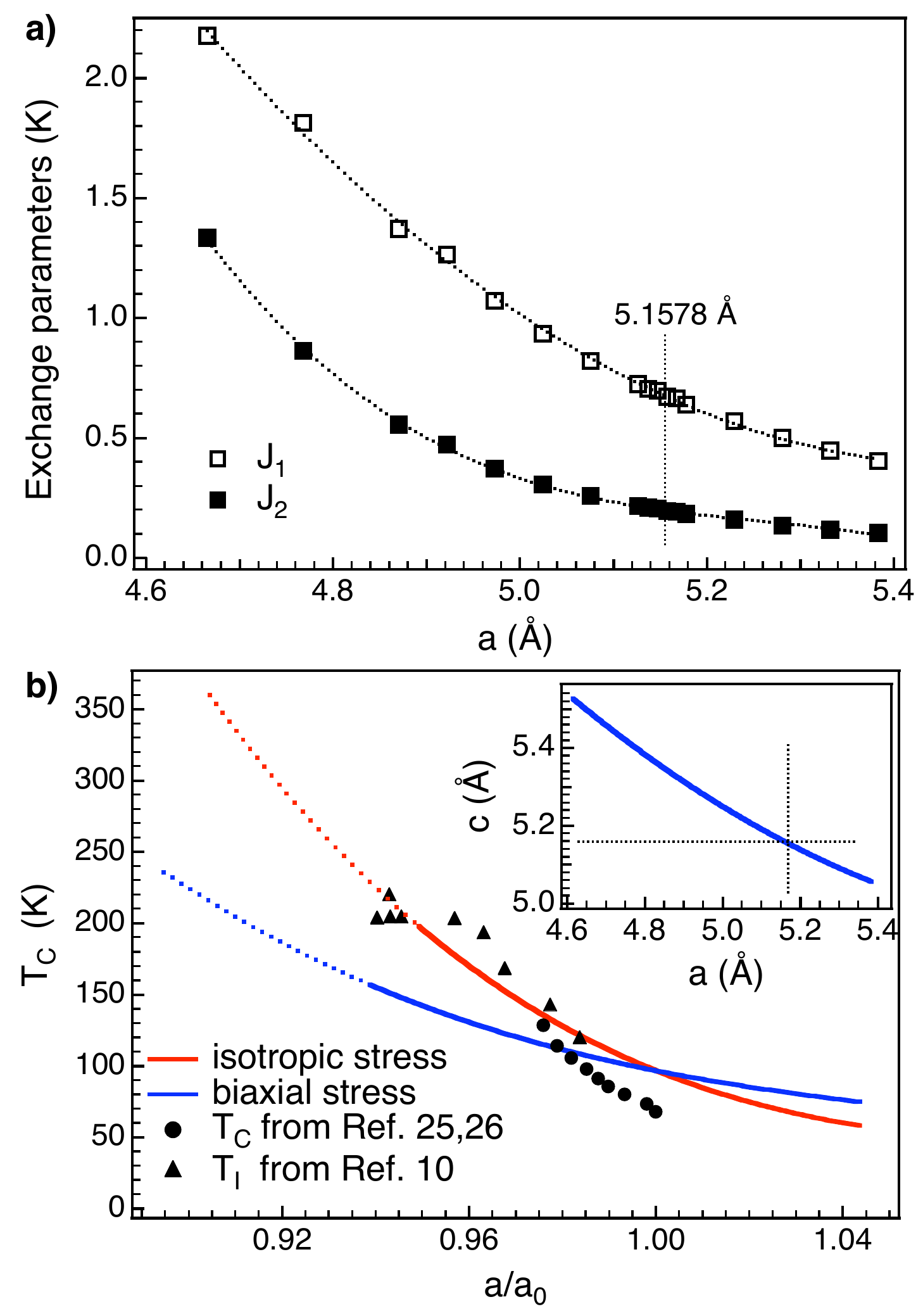}
\caption {(a)  The nearest neighbor and next nearest neighbor exchange constants, $J_1$ and $J_2$, as a function of lattice parameter for the isotropic stress case.    (b) The calculated mean field $T_c$ as a function of in-plane lattice parameter for the isotropic stress case and the biaxial stress state of an epitaxial film.  Both solid lines change to dashed lines at their respective IMT.    Included in the graph is the experimental $T_c$ -- or T$_I$ which tracks $T_c$\cite{dimarzio-prb-87} -- as a function of lattice parameters compiled from Ref \onlinecite{zimmer-prb-84}, \onlinecite{mcwhan-pr-66} (circles) and \onlinecite{dimarzio-prb-87} (triangles).  The inset shows the change in the out-of-plane lattice parameters in the biaxial stress case. }
\label{hydro-biaxial-lattice}
\end{figure}

In the isotropic stress case, we can extract $J_1$ and $J_2$  coupling constants from our calculations by relating the total energies of the various magnetic configurations (FM, AFM$_I$, AFM$_{II}$)\cite{smart-book} from the LSDA calculations with the Heisenberg model.  We assume that the third nearest neighbor coupling, $J_3$, is very small\cite{duan-prl-05,larson-jpcm-06}.  From these values, shown as a function of lattice parameter in Figure \ref{hydro-biaxial-lattice}a, we can calculate a mean field value for $T_c= (2/3)S(S+1)(12J_1+6J_2)$.   Our exchange parameters for the optimized bulk lattice parameter ($J_1=0.66$, $J_2=0.19$) are in good agreement with neutron measurements, and previous calculations.\cite{passell-prb-76,larson-jpcm-06}  

Previous experimental studies have looked at the effect of hydrostatic pressure on $T_c$\cite{mcwhan-pr-66,dimarzio-prb-87} and the unit cell volume\cite{zimmer-prb-84} in EuO.  In Figure \ref{hydro-biaxial-lattice}b we combine that data to plot the experimental change in $T_c$ as a function of lattice parameter.  Included in this figure is our calculation of the mean field $T_c$ of the isotropic case as the lattice parameter is changed.  There is quite good agreement between the experimental and calculated mean field $T_c$, with the expected over estimation of the mean field value. 

The calculations for the isotropic stress case indicate the closing of the Eu $4f$-Eu $5d6s$ gap, or IMT, at $a/a_0 = 0.95$.  This compares well with the appearance of a drude-like peak in the optical reflectivity and a jump in the room temperature resistivity at a pressure of $\approx130$ kbar, or $a/a_0 = 0.96$.\cite{zimmer-prb-84,dimarzio-prb-87} As seen in Figure 2, a saturation of the experimental parameter $T_I$, which is thought to track $T_c$, is also found around $a/a_0 = 0.96$.  If $T_I$ does in fact track $T_c$, then this saturation may be connected to the filled spin up $4f$  and empty $5d$ bands starting to overlap causing a mixed configuration of the J=7/2 magnetic Eu $4f^7$ configuration and the J=0 non-magnetic Eu $4f^6$ configuration.\cite{zimmer-prb-84}  Within our calculations for EuO with lattice parameters smaller than the IMT values, a J=7/2 magnetic configuration is always generated as the real many-body nature of the J=0 singlet state is not accessible from the LDA framework.  Therefore, our calculations do not include the possible disruption of the magnetic order from the $f^6$ configuration and we find $T_c$ continuing to rapidly rise after the gap has closed.  As the material becomes more metallic with increasing doping, other exchange mechanisms such as RKKY would start playing an increasingly important role.   The transition, however, will be smooth as the Fermi wave vector, $k_f$, in the metallic state will be very small due to the low doping level.

The mean field $T_c$ for the biaxial stress case is also included in Figure \ref{hydro-biaxial-lattice}b.  This is calculated as $T_c=(2/3)S(S+1)((4J_1^{\parallel}+8J_1^{\perp})+(4J_2^{\parallel}+2J_2^{\perp}))$.  The in-plane ($\parallel$) and out-of-plane ($\perp$) nearest and next-nearest neighbor distances are calculated from the optimized FM tetragonal structure at a given $a$ lattice parameter, and the appropriate $J_1$ and $J_2$ exchange parameter values are determined for these distances from the data presented in Figure \ref{hydro-biaxial-lattice}a.  The biaxial stress shows a smaller effect on $T_c$ for a given in-plane lattice parameter change as compared to the isotropic stress case.   This occurs because the average atomic distances change more slowly in the biaxial stress state due to the ability of the out-of-plane lattice spacing to compensate for changes in the in-plane lattice parameter.  The IMT is now found to occur at $a/a_0=0.94$.  As in the isotropic stress calculation, we find $T_c$ continuing to increase as lattice parameters are decreased beyond the IMT.  

The compensation for the in-plane lattice parameter change in the biaxial stress case is shown in the inset of Figure \ref{hydro-biaxial-lattice}a.  The relationship between the in-plane, $a$, and out-of-plane, $c$, lattice parameters for the biaxial stress state is given by $\frac{\Delta a}{a} = \frac{-2\nu}{1-\nu}\frac{\Delta c}{c}$, where $\nu$ is poisson's ratio.  $\nu$ is calculated to be ~0.15.  This is very close to the experimental value of 0.2 for NiO,\cite{james-ss-99} and the calculated value of 0.2 for GdN.\cite{duan-prl-05}

Magnetism in EuO is related to three main types of exchange mechanisms: local, or on-site exchange between the Eu $4f$ and other Eu orbitals; a Kramers-Anderson superexchange\cite{anderson-pr-1950}; and a number of other virtual exchange possibilities across the semi-conducting gap.  As expected we find that on-site exchange between Eu 4$f$ and 5$d$ is significant.  It will not, by itself, generate long range FM order, but plays a major role in the various virtual exchange mechanisms.  Another conceptually important mechanism is the Kramers-Anderson superexchange caused by the overlap of Eu 4$f$ and O 2$p$ orbitals. However, since the O $2p$-Eu $4f$ hopping parameters are quite small, as compared to O $2p$-$3d$ hoping in transition metal oxides, this antiferomagnetic component of $J_2$ plays only a minor role in EuO.  If  the gap between the O $2p$-Eu $4f$ states were to drastically reduce, increasing the overlap in energy of the states, this mechanism would become much more pronounced.  Nonetheless, this can only be achieved by changing the chemistry of the europium chalcogenides; lattice parameter changes are not enough.  Although the exact  decomposition of all possible virtual exchange mechanisms due to the excitations across the semiconducting gap is outside the scope of this work, by using a constrained potential in LSDA, one can change the energy of a specific orbital and effectively control the hopping integrals associated with that orbital.  This helps to separate out what types of changes in the band structure are most significant to the ferromagnetism. 

We find that in order to most effectively increase $T_c$, the hybridization between Eu $4f$ and $5d$ should be increased, as should the that of the O $2p$ and Eu $5d$ bands, while at the same time minimizing the O $2p$ to Eu $4f$ hybridization.  Since the Eu $4f$ band sits between the O $2p$ and the Eu $5d$ in EuO it is not possible to do all three at the same time.  However, the gap between the Eu $4f$ and $5d$ bands closes much more rapidly than the O $2p$ and Eu $4f$ band gap, so there is minimal lose in the ferromagnetic exchange due to the O $2p$-Eu $4f$ superexchange from decreasing lattice parameters.   Therefore, we find that decreasing lattice parameters in EuO is an ideal way of strengthening the ferromagnetic exchange.

\begin{figure}
\includegraphics[width=0.47\textwidth]{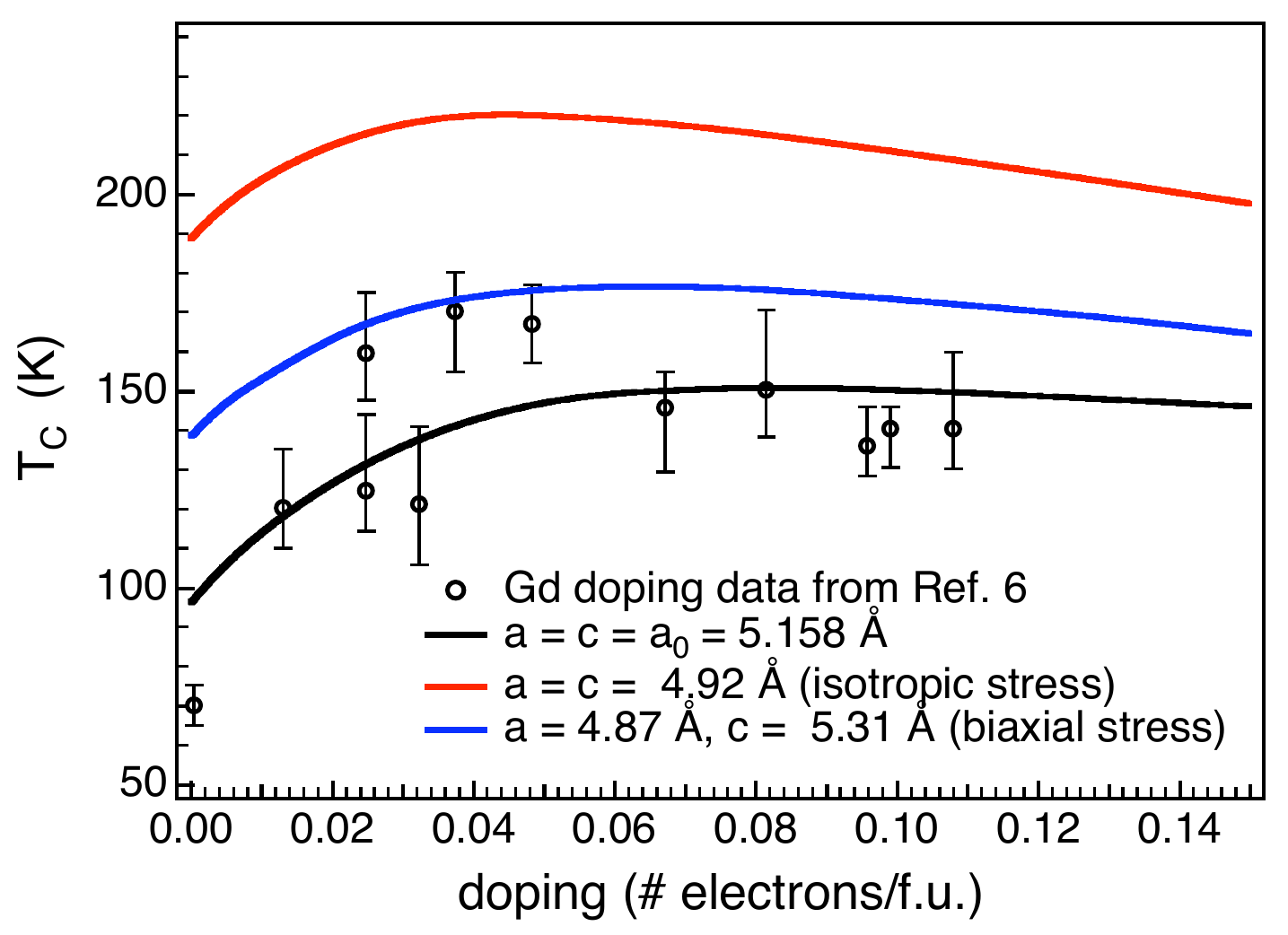}
\caption {The mean field $T_c$ as a function of electron doping for bulk ($a=c=a_0=5.158$\AA), isotropic stress ($a=c=4.92$\AA) and biaxial stress ($a=4.87$\AA, $c=5.31$\AA).  $T_c$ of Gd doping from Ref. \onlinecite{ott-prb-06}.}
\label{biaxial-DOS}
\end{figure}

Finally, by looking at the total energy changes due to a rigid band shift of the chemical potential for the FM, AFM$_I$ and AFM$_{II}$ configurations, we can also study the combined effect of isotropic or biaxial stress and electron doping on the mean field $T_c$.  Figure \ref{biaxial-DOS} shows these results for three situations; bulk lattice parameters, an isotropic stress case on the insulating side of the IMT ($a/a_0 = 0.955$), a biaxial stress case on the insulating side of the IMT ($a/a_0= 0.945$). At low doping, $T_c$ immediately increases in all cases because of the gain in total energy of the FM configuration relative to either of the AFM configurations. As doping increases, the energy cost to add extra electrons into both FM and AFM configurations becomes similar and a maximum in $T_c$ is reached.  Our calculations suggest that the DOS of a 5.5\% biaxially compressed epitaxial film would allow a significant increase in $T_c$, from 138 K to 175K.

Also included in Figure \ref{biaxial-DOS}, for comparison, are data from Gd doping experiments of Ott {\em et al}.\cite{ott-prb-06} In essence, Gd doping is adding one electron per Gd atom to EuO while leaving the S=7/2 spin configuration unchanged.   The maximum change in $T_c$ with Gd doping is about twice the size expected for electron doping via a rigid band shift of the chemical potential.  This suggests that there are some further effects present.  One possibility is that subtle changes occur in the $5d6s$ bands which effect the $4f$ to $5d$ exchange.   The presence of the smaller Gd 3+ ion may also provide a positive chemical pressure on the EuO matrix and decrease lattice parameters, activating all the previously discussed mechanisms.  It is worth noting that if the impressive $T_c$ enhancement from Gd is related to chemical pressure, then Gd would be a less effective means of increasing the $T_c$ of a film that is already biaxially compressed. 

In conclusion, we find that epitaxial influence can be used to increase the $T_c$ of EuO.  However, we show that the effect of the biaxial stress generated by epitaxy is not as effective as hydrostatic pressure due to the ability of the out-of-plane lattice parameter to partially compensate for the in-plane lattice changes.   The amount of out-of-plane compensation is minimized for the three-dimensional isotropic rock-salt-type structure in EuO because of its small poisson's ratio.   A material with a two-dimensional electronic structure would allow for the ultimate minimization of the out-of-plane compensation inherent in the biaxial stress state of epitaxy, although the likelihood of buckling or other structural modifications would be enhanced.  From the band structure perspective we find that the location of the near $E_f$ bands suggests that a decrease in lattice parameters is an ideal means of increasing $T_c$, as it strongly enhances the main ferromagnetic exchange mechanisms.  The closing of the semi-conductor gap is found to occur at a 5\%  lattice reduction in the isotropic stress case, or 6\% reduction in the biaxial stress state.   Although the closing of the gap will change the next nearest neighbor exchange mechanism, there is no sudden change in the magnetic properties.  Finally, we find that electron doping will lead to an increase in $T_c$ for isotropically and biaxially compressed films.

We gratefully acknowledge George Sawatzky for enlightening discussions.  The work was supported by the Natural Sciences and Engineering Research Council of Canada and enabled by the use of WestGrid computing resources.


\begin{thebibliography}{28}
\bibitem[{com()}]{comment_on_para_vs_AFM}
\bibinfo{howpublished}{The lack of long range order in the magnetic structure
  of the paramagnetic state means it can not be calculated within DFT.
  Furthermore, a non-spin polarized calculation will generate a different total
  energy due to the absence of local moments.}
\end{thebibliography}
\end{document}